\documentclass[11pt, a4paper, oneside]{article}

\usepackage{epsfig}
\title {Virtual Kathakali: Gesture-Driven Metamorphosis}
\author {Soumyadeep Paul, Sudipta N. Sinha, and Amitabha Mukerjee\\
Center for Robotics\\
Indian Institute of Technology, Kanpur \\
{\tt \small e-mail: \{spaul,snsinha,amit\}@iitk.ac.in} }

\begin{document}
\date{}

\maketitle
\begin{abstract}

Training in motor skills such as athletics, dance, or gymnastics is
not possible today except in the direct presence of the
coach/instructor.  This paper describes a computer vision based
gesture recognition system which is used to metamorphose the user into
a Virtual person, e.g. as a Kathakali dancer, which is graphically
recreated at a near or distant location. Thus this can be seen by an
off-site coach using low-bandwidth joint-motion data which permits
real-time animation.  The metamorphosis involves altering the
appearance and identity of the user and also creating a specific
environment possibly in interaction with other virtual creatures.
Unlike previous approaches to gesture based virtual reality, here i) a
single monochrome camera is used to track the arms, and ii) colour
differences are used to disambiguate situations where the arm is
occluding the user's body.

A robust vision module is used to identify the user, based on very
simple binary image processing in real time which also manages to
resolve self occlusion, correct for clothing/colour and other
variations among users. 
Gestures are identified by locating key points at the shoulder, elbow
and wrist joint, which are then recreated in an articulated humanoid
model, which in this instance, represents a Kathakali dancer in
elaborate traditional dress.  Unlike glove based or other gesture and
movement tracking systems, this application requires the user to wear
no hardware devices and is aimed at making gesture tracking simpler,
cheaper, and more user friendly. 

\end{abstract}

\section{Introduction}

Recognizing and tracking human gestures, with its great
promise for simplifying the man-machine interface, has seen
considerable emphasis in recent years. The main approaches have been
to use dedicated hardware such as dataglove or
polhemus sensors,
\cite{Pook/Ballard:1994,Sturman/Zeltzer:1994} or
visual recognition which requires little hardware but yields less
direct results 
\cite{	
Hunter/Schlenzig/Jain:1995,
Kortenkamp_et_al:1996,
Mishra/Singh/Prasannaa_et_al:1996,
Mukerjee/Dash:1998,
Mukerjee/Dash_et_al:1997,
Pavlovic/Sharma/Huang:1997,
Quek:1994,	
Rehg/Kanade:1994,	
Starner/Pentland:1995,
Wilson/Bobick:1998,
Wren/Azarbayejani_et_al:1997}.

One application of gesture recognition that is beginning to emerge is
off-site training for motor skills, e.g. in activities such as
athletics, surgery, theater/dancing, or gymnastics. Here the user's
motions can be transmitted using some low-bandwidth representation
(e.g. joint angles or facial expressions), and the instructor or coach at
the remote site can visualize the student using a local graphics model
at real-time animation rate, and provide appropriate feedback,
possibly illustrating the correct procedure through a similar virtual
metamorphosis channel.  For
example, a renowned master in the Kathakali dance form
\footnote{{\em Kathakali:} This celebrated dance tradition 
involves wearing elaborate costumes
and headgear, and also the use of special eye-masking paints and other
cosmetics.} 
may be able to
provide personalized feedback to a student far away - the sensation of
being co-located in the same virtual space makes communication much
more natural.  Furthermore, the master (or disciple) has the ability
to zoom in on a particular part of the performance or view the scene
from a particular vantage point, or to have the
actions repeated in slower speeds.  

\begin{figure} [hbtp]
\centerline{\psfig{file=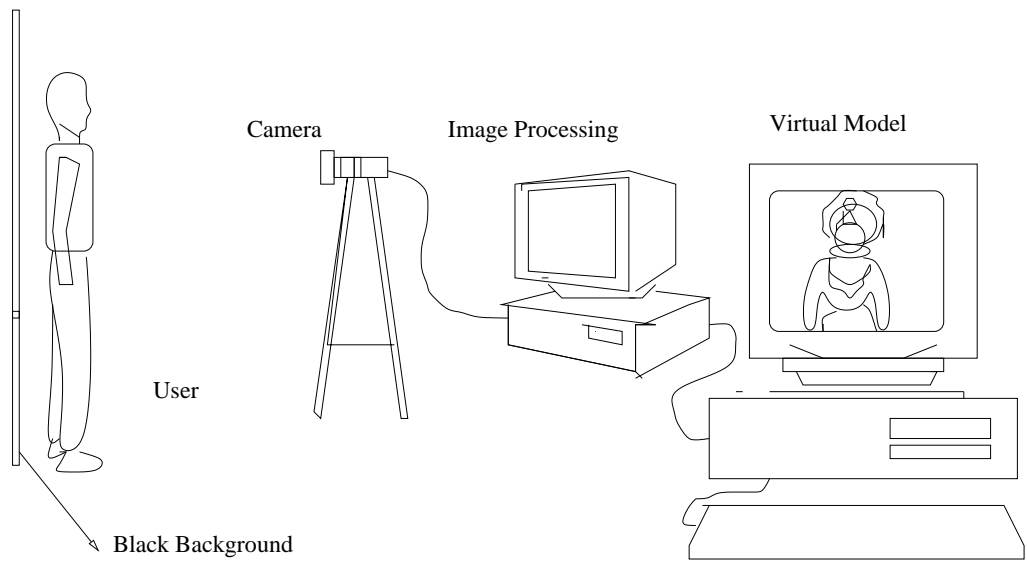,width=\columnwidth}}
\caption{{\em Virtual Modeling of User Motions.} The user moving in 
front of the camera sees himself as a Kathakali dancer in the Virtual
Model. The gestures of the user are transmitted to an articulated 
graphics model which then recreates it with the appropriate costume
and other embellishments.  This low-bandwidth data may be
used to create an off-site display
for obtaining  the trainer's feedback. }
\label{fig:flow}
\vspace{5mm}
\end{figure}

Of course, other usual Virtual Reality
applications such as full body interaction in a virtual space, as in
games or advanced chat rooms, can also be conducted with such a
system.  Figure
\ref{fig:flow} shows the basic setup that would be needed. 

Early approaches to gesture modeling used specialized arm motion
detection sensors \cite{Sturman/Zeltzer:1994}.  Such sensors encumber
the user and impose constraints on their motion to a certain extent.
The camera based model provides a simpler, more flexible, and far
cheaper alternative to other approaches. However, with the
camera the body pose is not directly available, and considerable
effort is needed in image processing.  Different parts of
this problem have been tackled for many years now:
\begin{itemize}
\item The DigitEyes system Rehg/Kanade 94 (\cite{Rehg/Kanade:1994}) 
recovered a detailed kinematic description of the hand 
using a 27 degrees of freedom full hand model and one or two cameras. 
\item Markov model based gesture identification 
\cite{Hunter/Schlenzig/Jain:1995,
Starner/Pentland:1995}.
\item Body tracking and behaviour interpretation
\cite{Wren/Azarbayejani_et_al:1997}.
\end{itemize}
In general, camera based systems are not able to simultaneously
identify both fine and gross motions since a
full body field of view reduces the accuracy available
for looking at the hand.  See
\cite{Pavlovic/Sharma/Huang:1997} for a recent survey of the field. 

\begin{figure} [hbtp]
\begin{center}
\begin{tabular}{cc}
\psfig{file=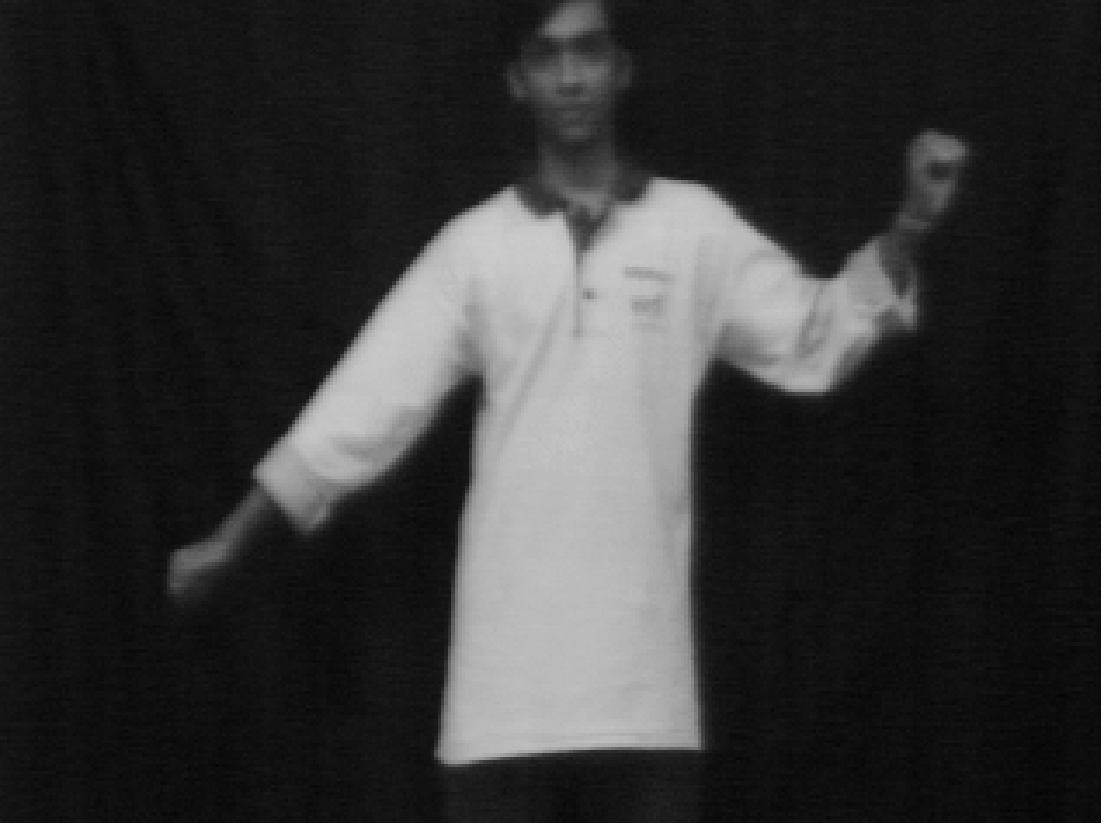,width=0.4\columnwidth} &
\psfig{file=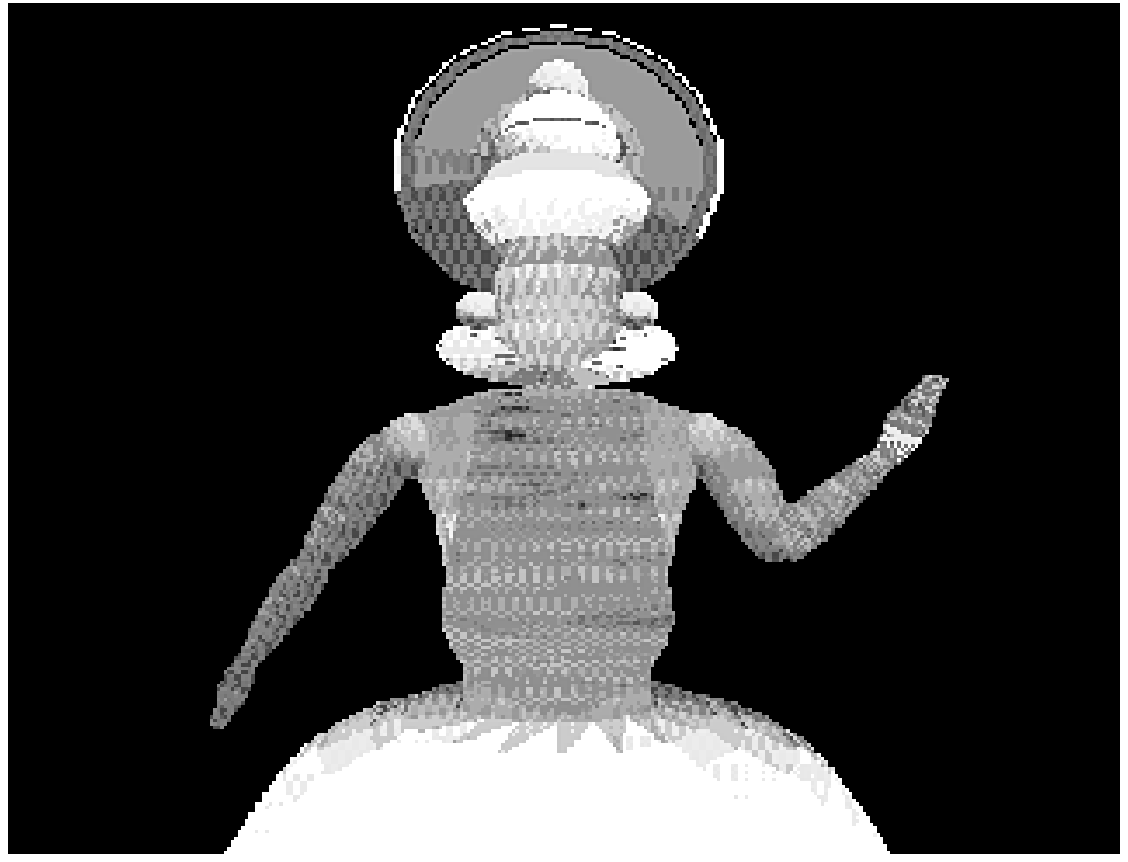,width=0.4\columnwidth} \\
\end{tabular}
\caption{{\em The User and the Model.} The User pose as seen by the
camera (from which the arm pose is to be detected), and the final
Kathakali Dancer Model as displayed to the user. }
\label{fig:virtual}
\end{center}
\end{figure}

Combining gesture recognition with graphics reconstruction provides
a virtual space where the user's action can be reflected. 
Applications in this genre include games
\cite{Mishra/Singh/Prasannaa_et_al:1996, Mukerjee/Dash_et_al:1997}, 
Virtual interaction spaces
\cite{Krueger:1991}, remote tele-operation
\cite{Mukerjee/Dash:1998}, and Virtual Metamorphosis
\cite{Ohya/Ebihara_et_al:1996}.
Our application is in the metamorphosis category
where the user is
metamorphosed as a Kathakali dancer in a virtual environment. 
The following section gives a brief outline of the techniques used in the 
paper.

\section{Outline}	

The system can be broken up into three modules:

\begin{itemize}
\item Real-time Detection of arm movements of the user
	(Section ~\ref{sec:armpose}).
\item Modelling of the Kathakali dancer
	(Section ~\ref{sec:modelling}).
\item Reproduction of the pose in the Kathakali dancer's model
	(Section ~\ref{sec:reproduce}).
\end{itemize}

Unlike other models that
use thermal imaging to obtain the user's silhouette
\cite{Ohya/Sengupta:1998}, the Virtual Kathakali 
system uses a visible-light monochrome camera against a black
background. The user's
silhouette is obtained by dynamically binarizing the images and the
3-D positions of the user's shoulder, elbow and the wrists are
obtained in real time from the image coordinates. 
This compact data is then transmitted to a
local or off-site virtualization system in real time.
Also, by using skin-tone colour/greyscale information, it is possible
to identify occlusion, which is not possible to do in thermal
imaging systems. 
The overall cost of this system is likely to be several times less
than that of other comparable systems used in Virtual Metamorphosis
systems. 

The next phase is to create
an articulated 3D model that will follow the user's poses
and reflect the traditional costumes of a classical Indian dance form
such as the Kathakali. 
The 3D model needs to have 
appropriate motion constraints at the joints and suitable
dress/headgear/texture.
The 3D arm pose sequence is now 
communicated to the graphics model, which recreates it as an animated
graphics display. 
In this process, 
the very low-bandwidth joint angle data can be used to animate the
3D Dancer model.

\section{Real Time detection of User's Arm Pose}
\label{sec:armpose}

In the initial pose, the user stands with his arms separated wide
apart and the following calibration data are obtained :-
\begin{itemize}
\item arm-length
\item range of pixel intensity within which the pixels corresponding to his 
body tone lie
\item arm width 
\item width of his shoulders
\end{itemize}

To simplify the image processing costs, the user is required
to stand in front of a dark background. This permits
image binarization based on a dynamic threshold,
set at a point of sharp variation in the intensity histogram. 
Noise may yet
interfere with the robust determination of arm posture, and 
Gaussian convolution is used to smooth out some of the noise.

\begin{figure}
\begin{center}
\begin{tabular}{cc}
\psfig{file=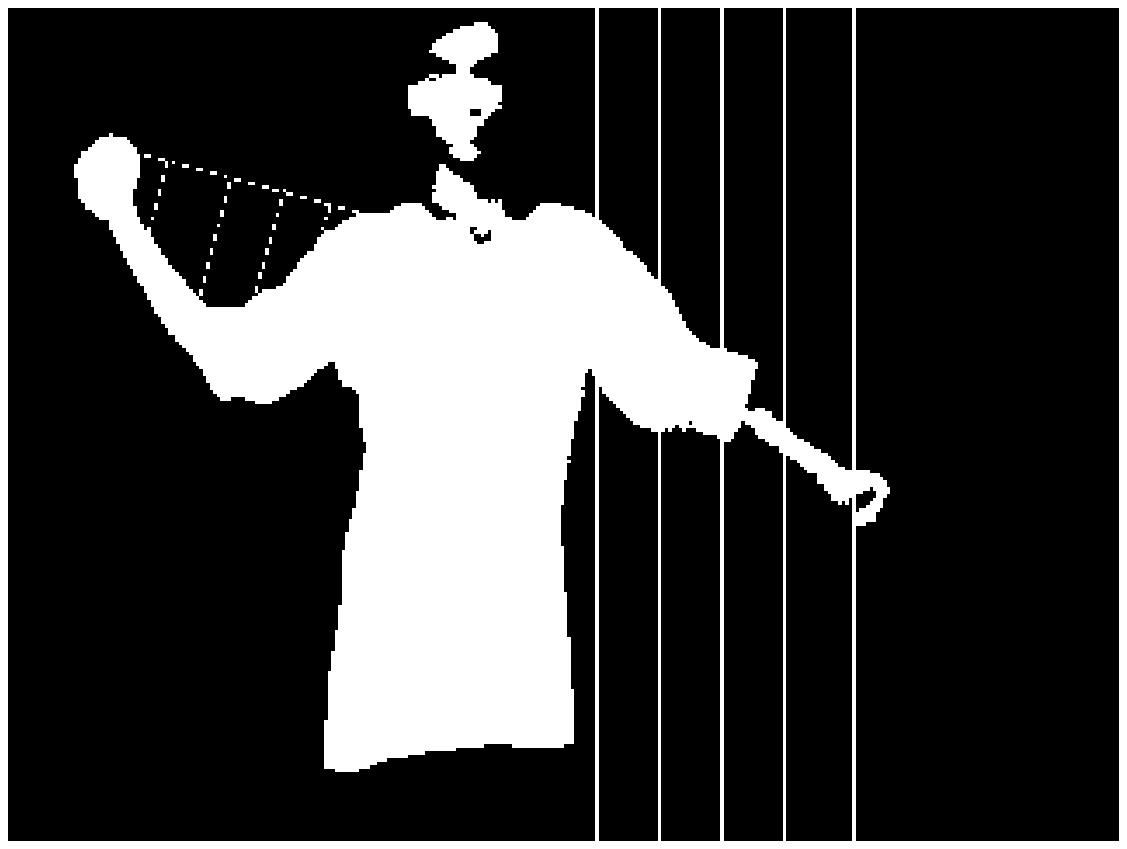,width=0.4\columnwidth} &
\psfig{file=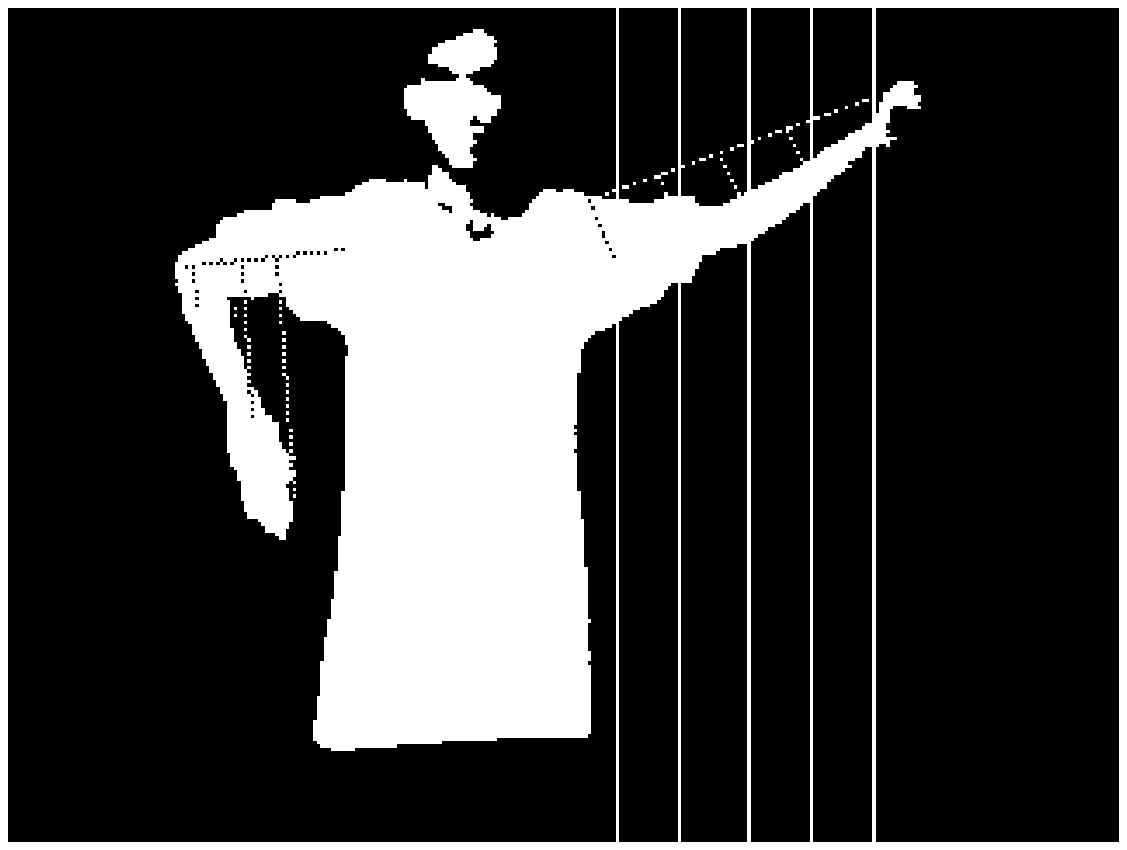,width=0.4\columnwidth} \\
\end{tabular}
\caption{{\em Image Processing Results}. Left: Locating outermost tip. 
Right: Finding the elbow and wrist joints.}
\label{fig:segments}
\end{center}
\end{figure}

The pose of the user's arms at each instant is obtained by identifying
the elbow and wrist in the image.  The body width is identified based
on points of high intensity change in the lower parts of the image.
The elbow and wrist are distinguished by different techniques
depending on whether the hand is occluding the body
(Section \ref{sec:occlusion_module}) or not. In the
latter case, a fast and simple
technique is to locate
the extreme points in the image and test if the line joining it to the
shoulder is part of the arm or not (Figure \ref{fig:segments}).
Based on this and the initial calibration information,
the 3D pose of the arm is estimated based on 
foreshortening. 

\begin{figure}
\begin{center}
\begin{tabular}{c}
\psfig{file=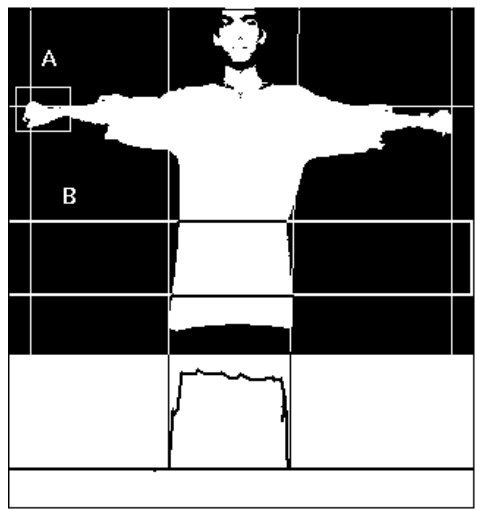,width=0.46\columnwidth} \\
\end{tabular}
\end{center}
\caption{{\em Calibration phase}. User stands with
arms apart. The body width, arm length and arm width are detected.
The histogram for the box B (shown below the image) is used to
detect body width, and a similar analysis in the region A
is used to obtain the skin
shade. 
}
\label{fig:initial}
\vspace{5mm}
\end{figure}

\begin{figure}
\begin{center}
\begin{tabular}{ccc}
\psfig{file=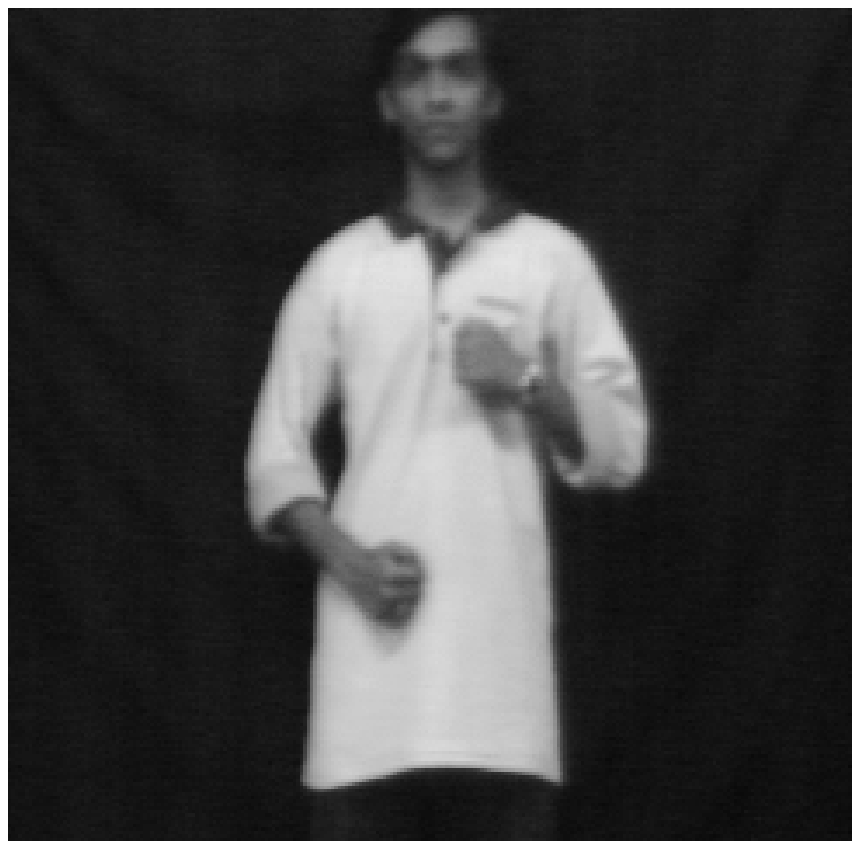,width=0.4\columnwidth} &
\psfig{file=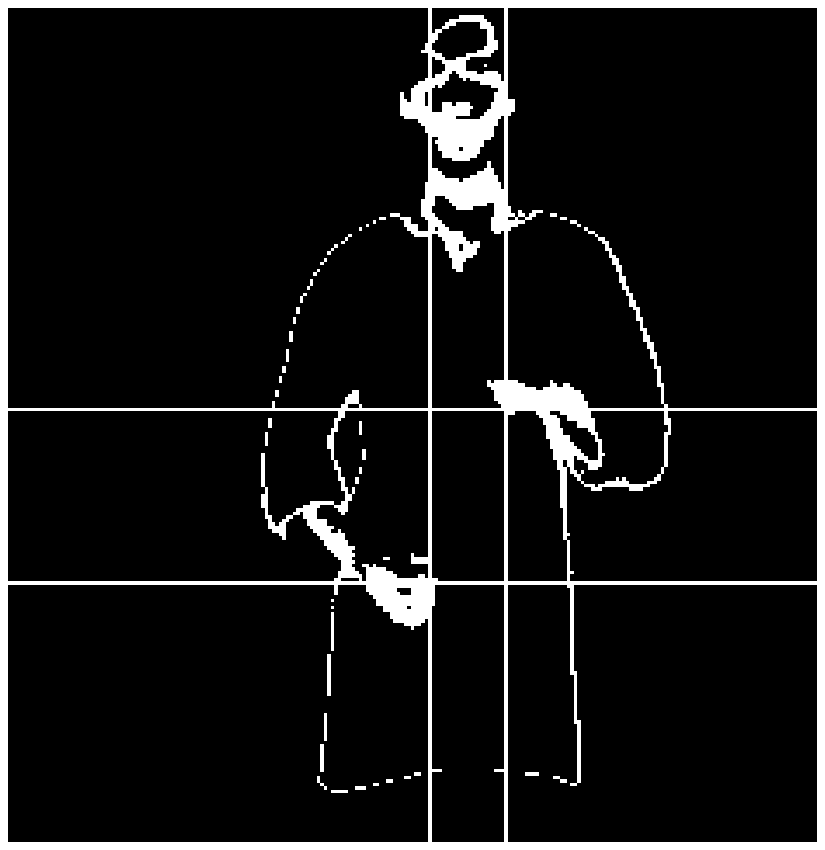,width=0.4\columnwidth} \\
\end{tabular}
\caption{{\em Resolving Occlusion}. 
The pixel intensity range for the user's skin colour / clothes are used to
distinguish parts of the arm that may be occluding the user's
body. }
\label{fig:occlusion}
\end{center}
\end{figure}

\subsection{Resolving Occlusion}
\label{sec:occlusion_module}

Many dance poses involve the hand being in front of the body; these
postures are particularly important in many {\em mudras}\footnote{{\em mudras:}
certain specific postures or positions in Indian dance forms.}. 
The binarized processing described above is not sufficient for
resolving this occlusion, so the wrist is identified in the image
based on the greyscale skin tones (see Figure~\ref{fig:initial}). 
During the
performance, for each grabbed image, the pixels outside this 
intensity range are discarded. If the user is wearing
clothes contrasting with his body color, only the pixels corresponding to his 
body parts retain the high value of intensity. This leads to an effective 
separation of the hand from the body in cases of self-occlusion (see 
Figure ~\ref{fig:occlusion}). The hand can now easily be detected.

In the case where the 
hand is before the body, we still need to identify the elbow for constructing 
the 3-D arm pose. 
This is simple here, since the 
outermost tip of the image corresponds to the elbow joint. 

\section{Modelling of Kathakali Dancer}
\label{sec:modelling}

The first step is to create the wireframe of the human
using primitives. 
A frustum with variable elliptical cross-sections has been used as a 
building block for modelling the Kathakali dancer.
The outer surface of this primitive is realised by a triangular mesh. 
A stack of such primitives are used to model the head, arms and body
separately. Though the resolution can be easily controlled, 
a very high resolution is sacrificed for faster rendering. 

The head has three degrees of freedom i.e.. 
it can be twisted, bent forward and
sideways. Moreover, the movements are limited in the range that is
humanely possible. 
To model the arms, first the shoulder
joints with three degrees of freedom have been created.  Next, the upper arm, 
the lower arm and the wrist are created using the frustum primitive. 
The elbow and wrist
joints are simulated as hinge joints i.e.. with only one degree of
freedom.   

The Kathakali dance form is famous for its elaborate apparel and
ornamentation. Hence, to give a realistic effect, texture mapping 
has been used to create an appropriate pattern on the front of the dress.
The image used for texture mapping has been obtained from images of
Kathakali dancers.

Various other Graphics rendering techniques like Lighting and Shading 
have been used to create the effect of a 3D Virtual World. The real time 
nature of the application requires the use of special techniques for 
optimizing performance.

\begin{figure}
\begin{center}
\begin{tabular}{cc}
\psfig{file=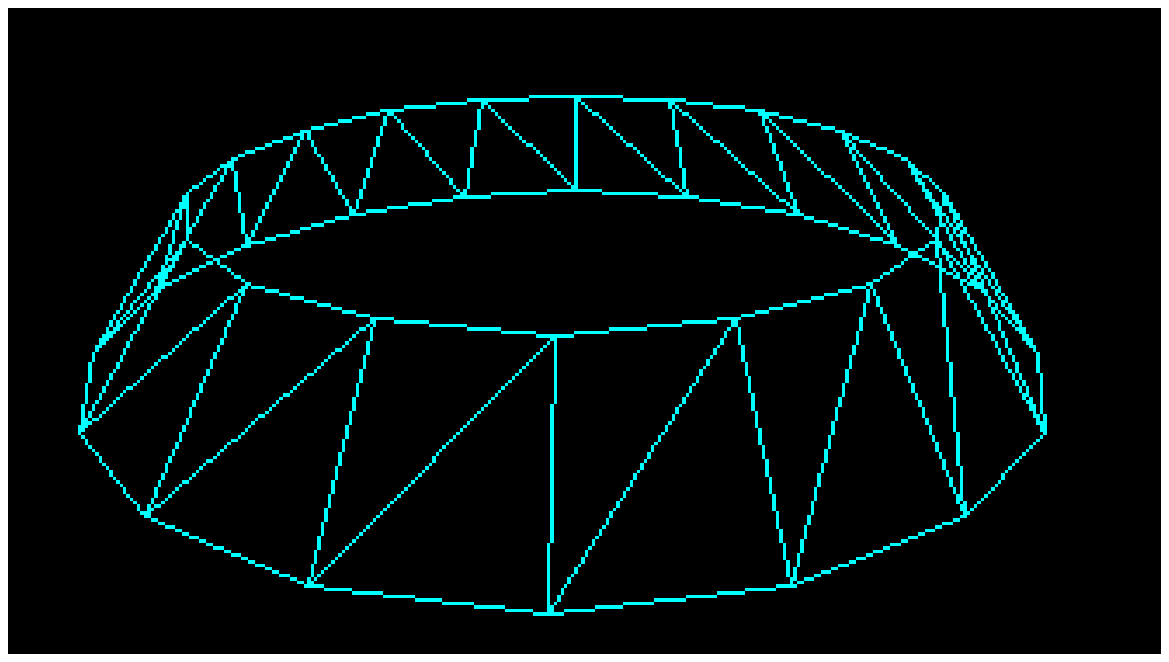,width=0.4\columnwidth} &
\psfig{file=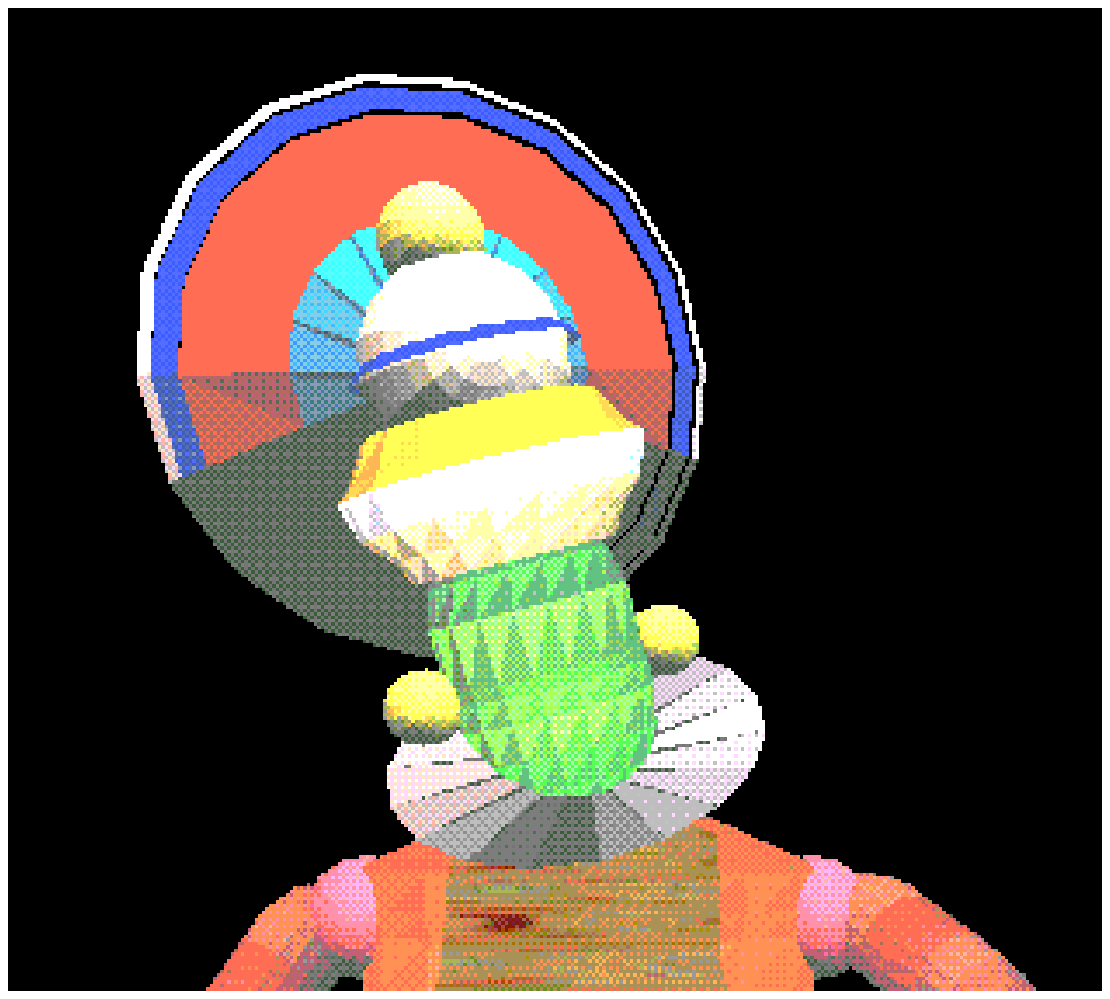,width=0.4\columnwidth} \\
\end{tabular}
\caption{{\em Graphics Modeling}. The frustum primitive 
has been used in several constructions, for example the headgear 
shown to the right.}
\label{fig:Graphics}
\end{center}
\end{figure}
 
\begin{table}
\begin{center}
\begin{tabular}{|c|c|c|} \hline
{\it Label} & {\it Joint description} & {\it Number of DOF} \\ \hline \hline
A & Neck joint & 3 \\  \hline
B,C & Shoulder Joints & 3 \\  \hline
D,E & Elbow Joints & 1 \\ \hline
F,G & Wrist Joint & 1 \\ \hline
\end{tabular}
\caption{{\em Modelling of Joints} This Table shows the
degrees-of-freedom
and constraints of the labelled joints in Figure~\ref{fig:Armtable}.}
\label{table:JointModel}
\end{center}
\end{table}

\begin{table}
\begin{center}
\begin{tabular}{|c|c|c|c|c|} \hline
{\it Head} & {\it Upper-arm} & {\it Forearm} & {\it Palm} & {\it Body,Dress} \\ \hline \hline
400 & 350 & 250 & 100 & 350  \\  \hline
\end{tabular}
\caption{{\em Number of Polygons} This Table shows the number of polygons used
to create the model of Figure~\ref{fig:Armtable}.}
\label{table:NumPoly}
\end{center}
\end{table}

\begin{figure}
\begin{center}
\begin{tabular}{c}
\psfig{file=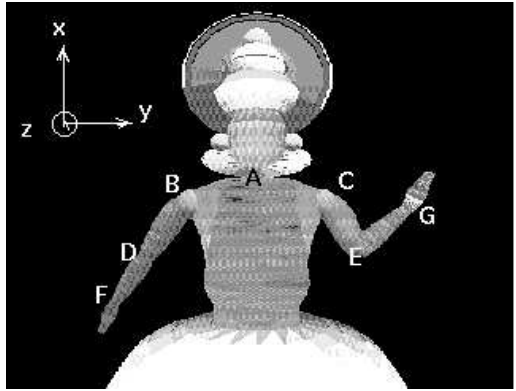,width=0.50\columnwidth} \\
\end{tabular}
\caption{{\em The Articulated Model}. The human upper body is modelled
as a set of rigid links connected by joints with different ranges of 
motion. The labels for the different joints are shown here
(see Table \ref{table:JointModel}). }
\label{fig:Armtable}
\end{center}
\end{figure}

\section{Reproduction of User's Pose in Virtual Model}
\label{sec:reproduce}

Based on the initial data about user's arm-length and forearm-length the
3D position can be recovered using the foreshortening seen in the
image. The angles of the upper arm with respect to 
the shoulders and the forearm with respect to the upper arm
are calculated and 
used to apply appropriate transformations of the limbs in the
virtual metamorphosis model.

This application was developed on a
a  PentiumII/Linux-OpenGL PC for the graphics
and an older  486 PC/DOS with a Matrox Image Processing card
under Matrox Imaging Library(MIL 2.1). 

\begin{figure}
\begin{center}
\begin{tabular}{|cccc|}\hline
\psfig{file=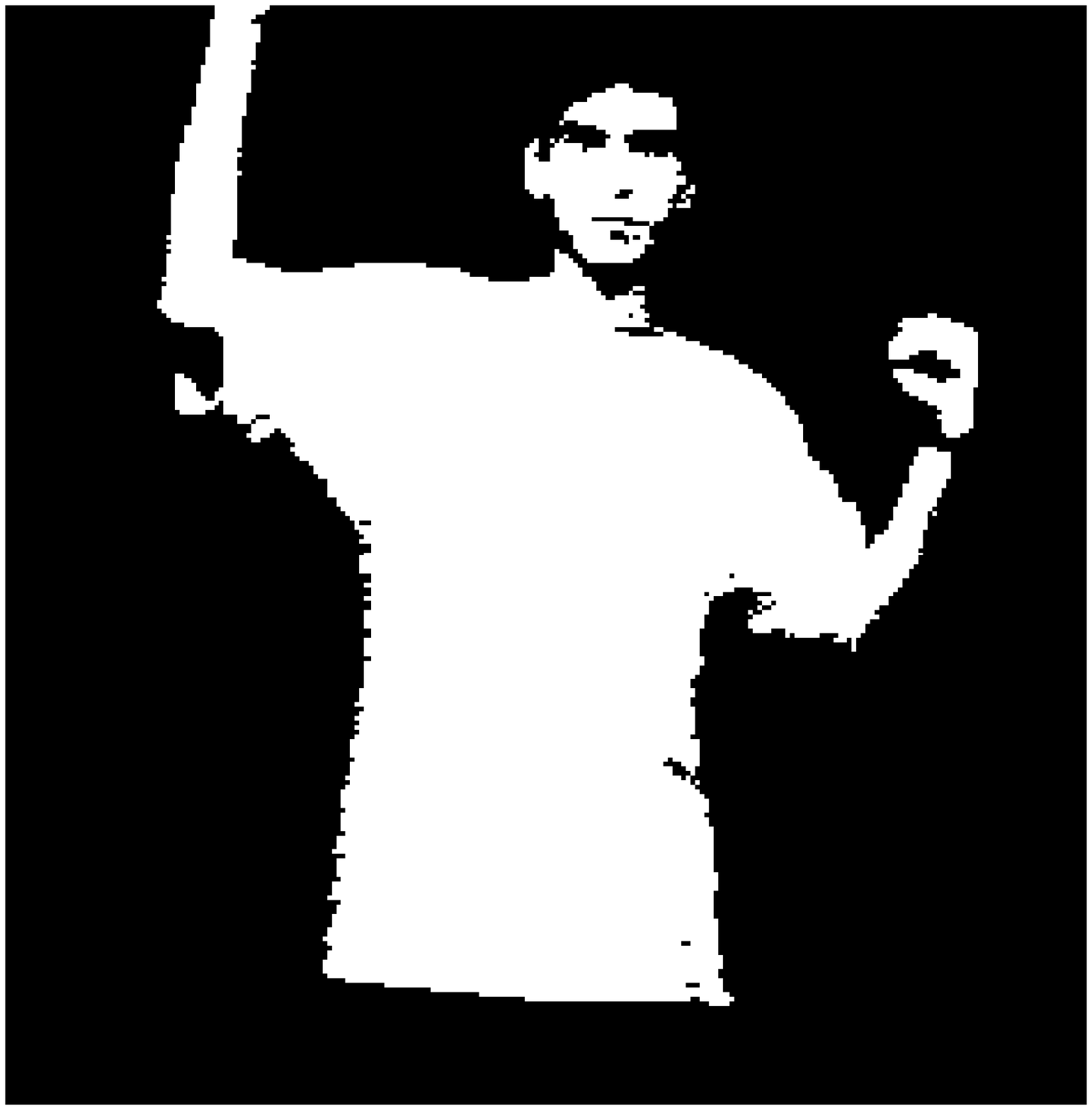,width=0.23\columnwidth}   & 
\psfig{file=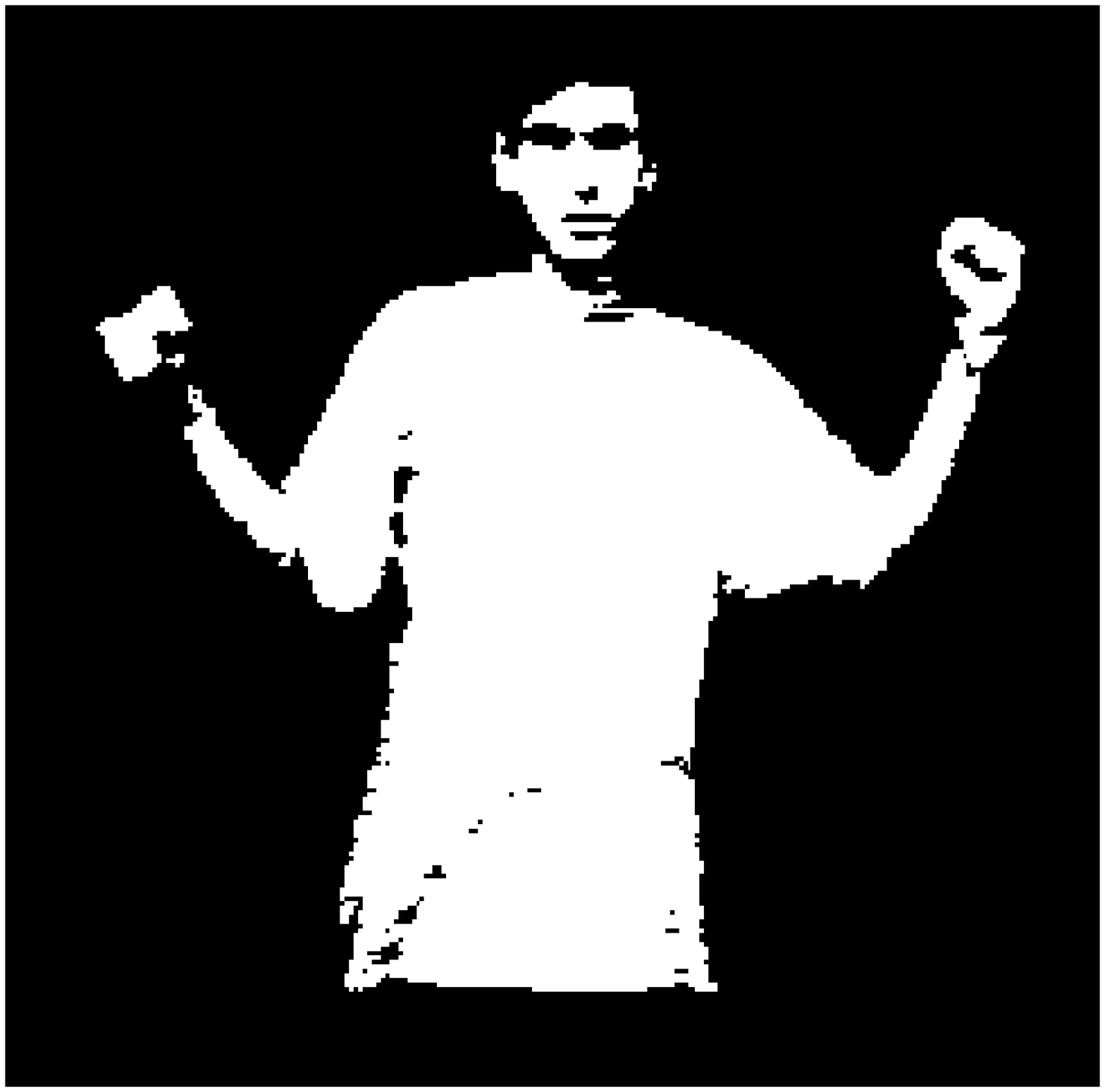,width=0.23\columnwidth}  &
\psfig{file=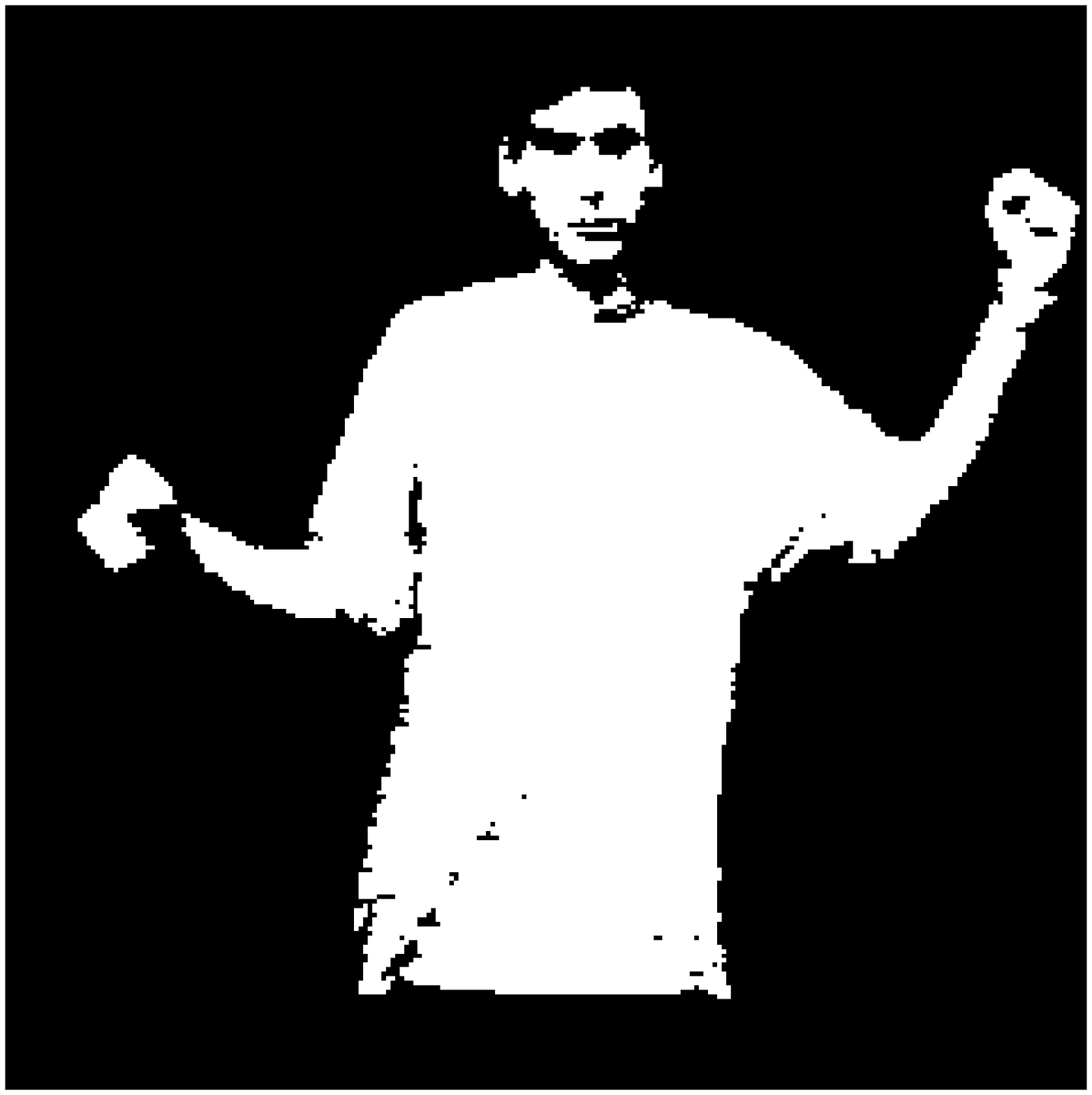,width=0.23\columnwidth}  &
\psfig{file=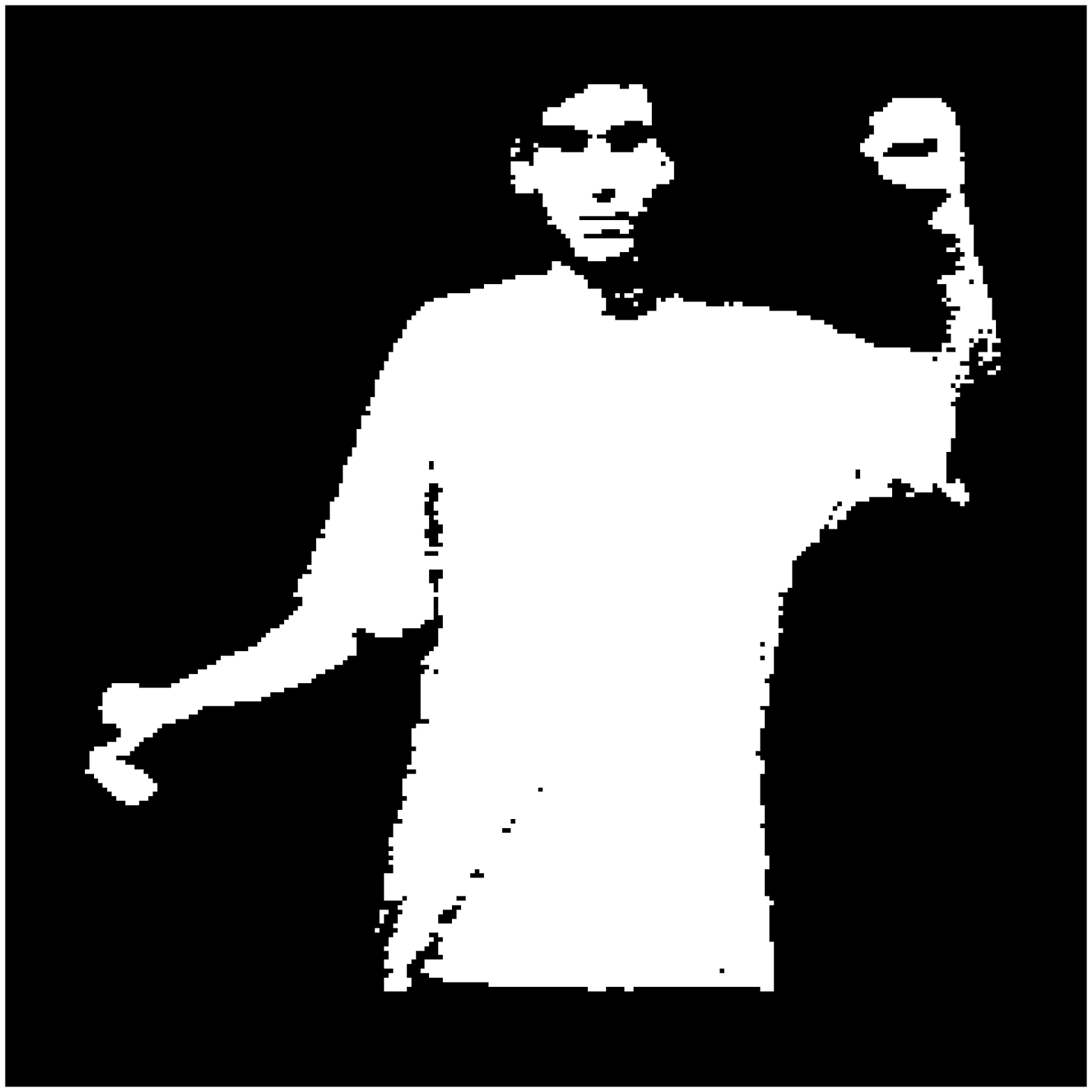,width=0.23\columnwidth}  \\ \hline
\psfig{file=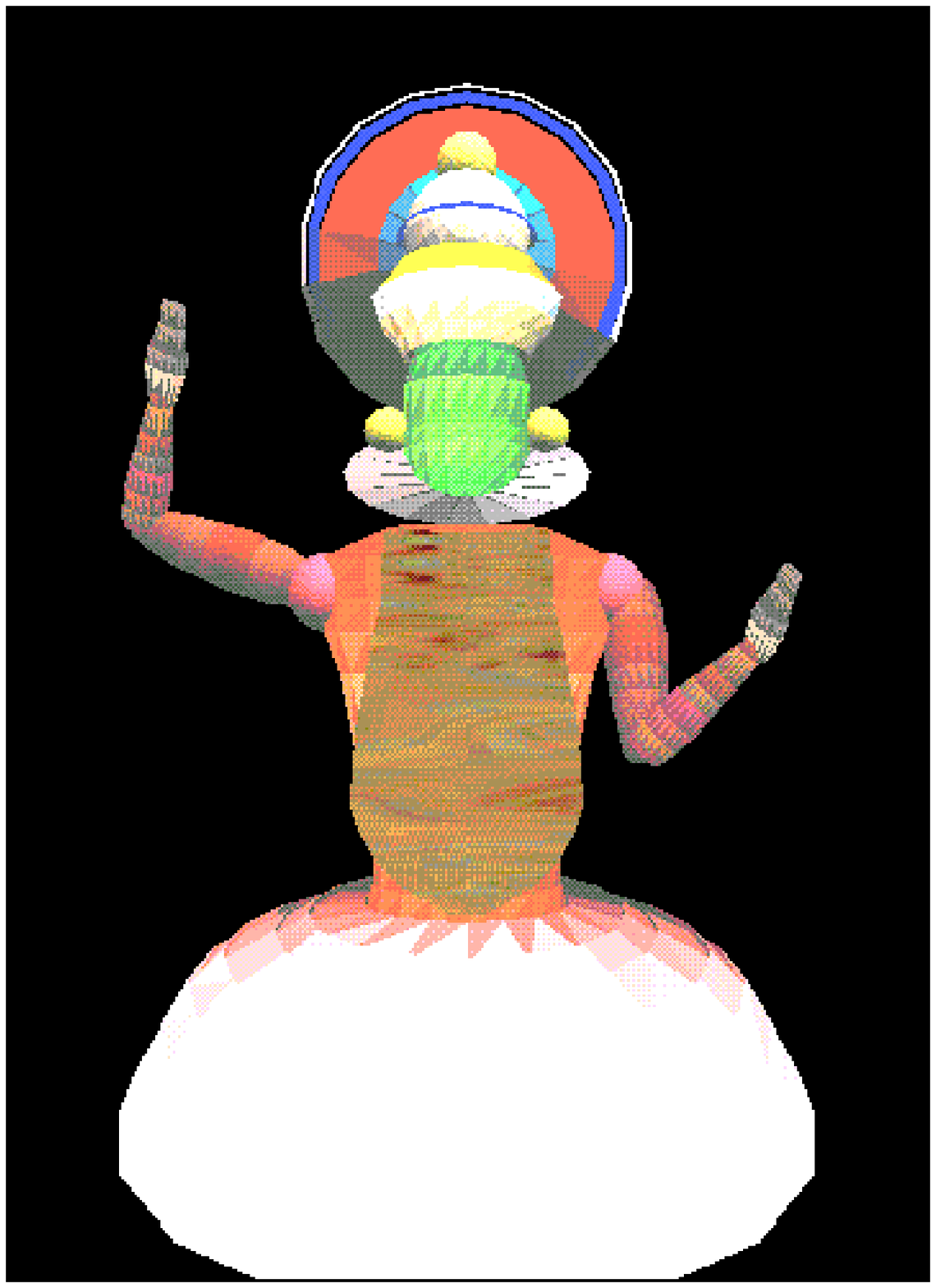,width=0.23\columnwidth} &
\psfig{file=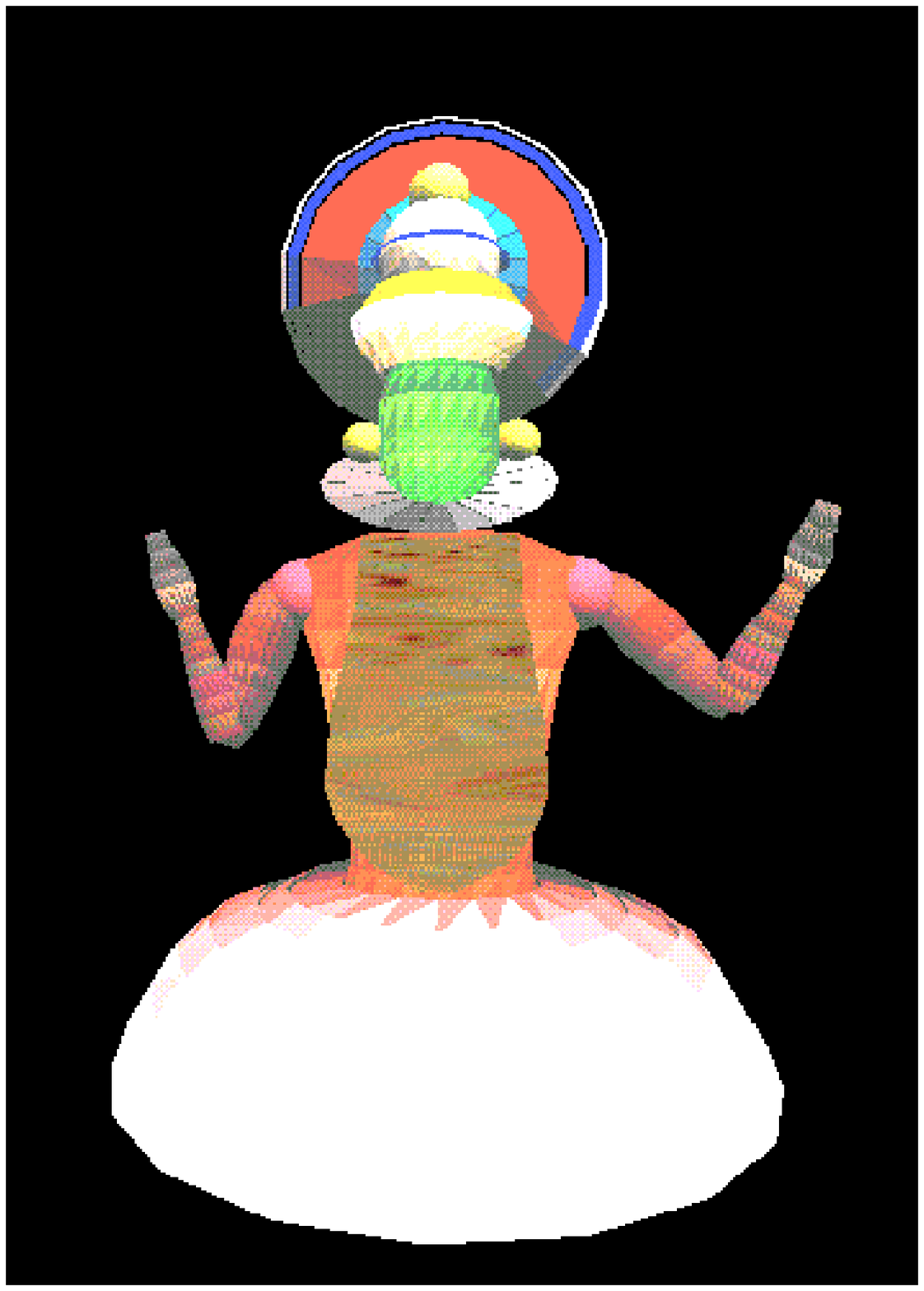,width=0.23\columnwidth} &
\psfig{file=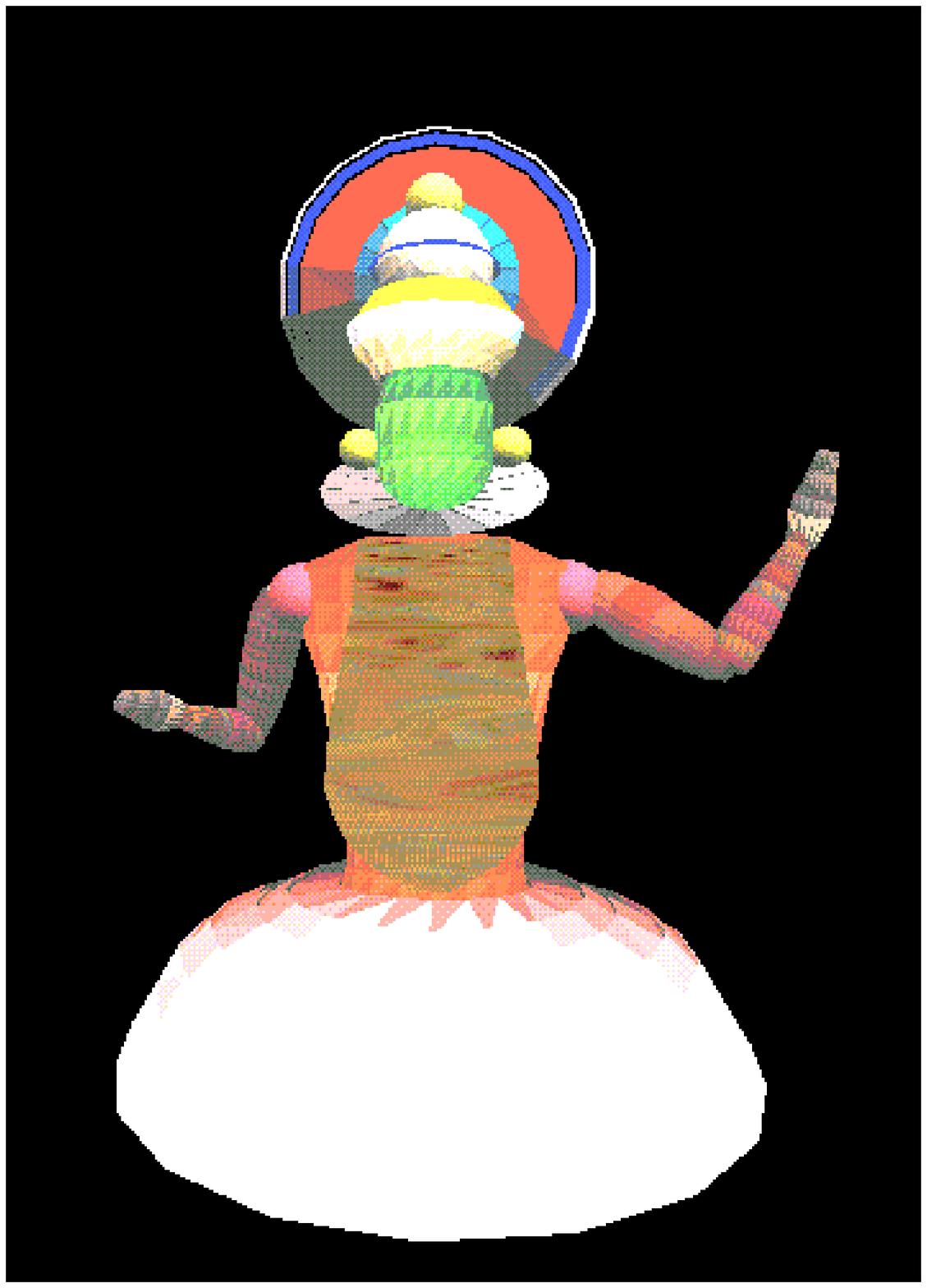,width=0.23\columnwidth} &
\psfig{file=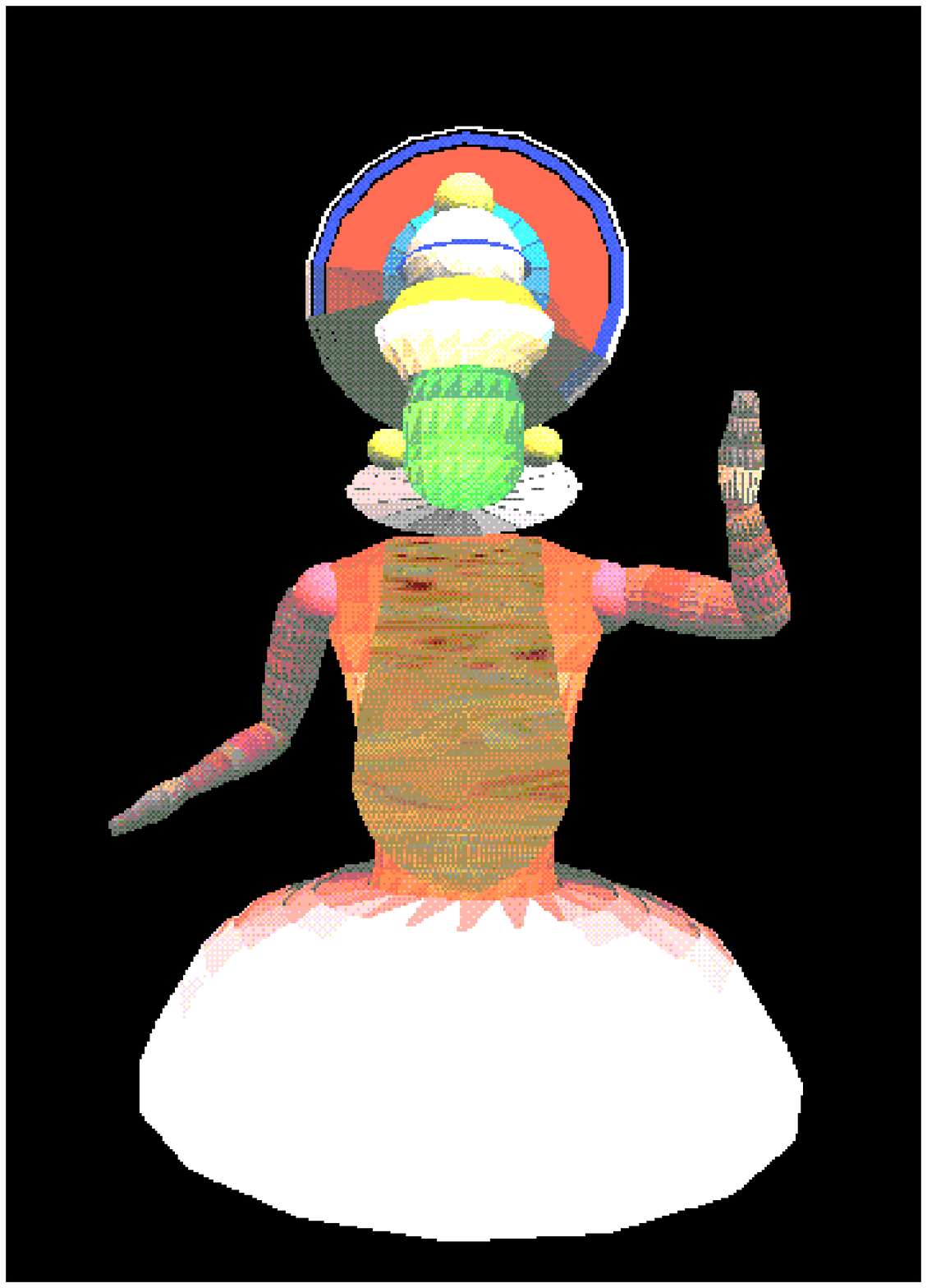,width=0.23\columnwidth} \\  \hline
\end{tabular}
\caption{{\em User's and the Model's poses}. 
Four frames representing a motion sequence are shown along with the 
graphics reconstruction (lower row). 
For each frame, twelve
joint angles are passed from the image interpretation module to the
graphic generation module. 
}
\label{fig:Output}
\end{center}
\end{figure}

Some images with user's poses and 
the corresponding graphics output are shown in Figure~\ref{fig:Output}. 

\section{Conclusions}

In this paper, we have presented a communication environment
in which a person could transform himself into any character he
desires in a virtual environment. In our case, the character was that
of a Kathakali Dancer. Computer Vision techniques are used for
passive detection of user's arm pose in real time. Some simple
constraints on the user (that he stand in front of a dark background,
and that his skin tone differs from his clothing) result in being able
to use some extremely simple and fast detection methods based on
dynamic threshold binarization. 
A single camera is
used and the user does not need to wear any special devices.
The 3-D model is instantiated during the calibration phase, and
the user's motions are reproduced using various transformations on the
dancer model. Our implementation showed good performance with a
processing speed of 5 frames per seconds on a very old and simple image
processing system. 

The primary application for such a system would be in the coaching of
motor skills with distant trainers and coaches. In the arts, such a
system would permit quick informal recording of creative insights,
which would otherwise require elaborate stage and makeup arrangements
before being presented in the final form. In general, the expansion of
Internet, even with low-bandwidth connectivity, will permit gestural
interaction of this type as long as major computational tasks such
as identification of limb postures and recreation/graphics output 
are performed on the local machines. In the coming decades, systems of
this type are likely to have a profound effect on the declining trend
in artistic traditions worldwide. Also, in sports as well as critical
operations such as surgery, this could enable expert trainers to provide
guidance to a far larger set of students than would be possible in a
face-to-face mode. 

Since the image processing is carried on an image in 2D, the system is
not able to resolve between two arm
poses with the same projection, as when the entire 
arm is on a horizontal plane. 
However, due to the very nature of this deficiency, it will not matter
to the viewer so long as it is viewed from the same angle. 

In this work, we have only used the arm pose of the user to control
the virtual model.  
In traditional dance form, facial expressions 
and finger movements constitute an important component of the dancer's
emotional expression ({\em abhinaya}). In the current phase, with a
single camera of fixed resolution, this is not possible; in fact, no
vision system today can model both the finger and the gross body
motions. However, some beginnings have been made towards integrating
face recognition by having a camera
look down on the user's face from a fixture mounted on the head itself. 
With multiple cameras, one camera could be used for the full-body
field of view, whereas one more other cameras could track the
important aspects of the dance - particularly the two hands and also
the face. Such a multi-scale imaging and tracking system would enable
detailed reconstructions of the important aspects of the scene. 

Moreover the system, which supports only a single user at present, can
easily be extended to support multiple users all sharing the same
virtual environment. Hence creation of Virtual Theaters become a
distinct possibility where different actors situated far away from
each other can be actors in the same virtual theater. Since the
bandwidth required by the system is very small, transmitting the data
over the Internet is a feasible option. Another challenging problem is
to have real and virtual actors share the same space in a seamless
manner from the audience's viewpoint. 
	
\small
\bibliographystyle{plain}
\bibliography{hci,frombib}

\end{document}